\documentclass[apl,reprint,twocolumn,superscriptaddress]{revtex4}
\usepackage{graphicx}
\usepackage{fancyhdr}
\usepackage{amsmath}
\usepackage[colorlinks=true, allcolors=blue]{hyperref}

\begin{document}

\preprint{...}

\title{Transient dynamics of strongly coupled spin vortex pairs: effects of anharmonicity and resonant excitation on inertial switching}

\author{E. Holmgren}
\affiliation{Royal Institute of Technology, 10691 Stockholm, Sweden}

\author{A. Bondarenko}
\affiliation{Royal Institute of Technology, 10691 Stockholm, Sweden}
\affiliation{Institute of Magnetism, National Academy of Science, 03142 Kiev, Ukraine}

\author{M. Persson}
\affiliation{Royal Institute of Technology, 10691 Stockholm, Sweden}

\author{B. A. Ivanov}
\affiliation{Institute of Magnetism, National Academy of Science, 03142 Kiev, Ukraine}
\affiliation{National University of Science and Technology MISiS, Moscow, 119049, Russian Federation}

\author{V. Korenivski}
\affiliation{Royal Institute of Technology, 10691 Stockholm, Sweden}

%\date{\today}

\begin{abstract}
Spin vortices in magnetic nanopillars are used as GHz oscillators, with frequency however essentially fixed in fabrication. We demonstrate a model system of a two-vortex nanopillar, in which the resonance frequency can be changed by an order of magnitude, without using high dc magnetic fields. The effect is due to switching between the two stable states of the vortex pair, which we show can be done with low-amplitude fields of sub-ns duration. We detail the relevant vortex-core dynamics and explain how field anharmonicity and phase control can be used to enhance the performance.

\end{abstract}

\maketitle

The dynamics of a spin vortex is well represented by the motion of its core -- a point-like object, about 10 nm in lateral size,\cite{Wachowiak2002,Shinjo2000} with the magnetization out of the plane,\cite{Feldtkeller1965} ``up" or ``down", referred to as the core polarization. The resonance mode of an isolated vortex is gyrotropic, where the core oscillates in a large-radius circular trajectory at a sub-GHz frequency.\cite{Voelkel1994,Guslienko2002} It has been shown that this mode of motion can be excited using either a high-frequency magnetic field\cite{Waeyenberge2006,Chou2007} or a spin-polarized current.\cite{Kasai2006,Moriya2008,Bolte2008} The gyrotropic resonance has been studied extensively, including direct observation using time-resolved X-ray imaging,\cite{Choe2004} optical Cotton-Mouton effect,\cite{Argyle1984} and MOKE.\cite{Park2003} Such variety of vortex dynamics with its low dissipation make spin vortices promising for applications in memory\cite{Pigeau2010} and rf oscillators\cite{Pribiag2007,Mistral2008}. Several multi-vortex systems have been explored for use in oscillators, including lateral vortex arrays\cite{Haenze2014} and vertically stacked vortex pairs.\cite{Wintz2016} 

Pairing of vortices can give rise to dynamic properties qualitatively different from those of the individual vortices constituting the pair, as has been shown in recent studies on lateral exchange-coupled vortex pairs\cite{Hata2014} as well as vertical magnetostatically-coupled symmetric\cite{Haenze2016} and asymmetric\cite{Lebrun2014,Stebliy2017} vortex pairs. The precise nature of the collective dynamics of a vortex pair depends on the strength of the inter-vortex interactions as well as the exact core/chirality configuration within the pair.

We previously showed that vertically stacked vortex pairs with strongly-coupled parallel (P) cores and antiparallel (AP) chiralities (P-AP pairs) possess a collective resonance mode of anti-phase rotation about the pair's "center-of-mass" at a frequency an order of magnitude higher than that for the gyration of an individual vortex core.\cite{Cherepov2012} A dc field of a few mT forces the cores apart, thereby creating a bistability of the coupled and decoupled core states, which at room temperature is limited to microseconds in lifetime.\cite{Bondarenko2017} 

In this letter we experimentally confirm that such a bistable state can be long-lived by conducting experiments at 77 K and demonstrate that a core-core pair decouples via nontrivial transient dynamics. The system exhibits inertial switching, amplified by phase-tuning the ac field on the scale of one half the period of the rotational resonance. We also show how anharmonic excitation can be used to enhance the switching speed and reliability.

\begin{figure*}[!t]
\includegraphics[width=6.5in]{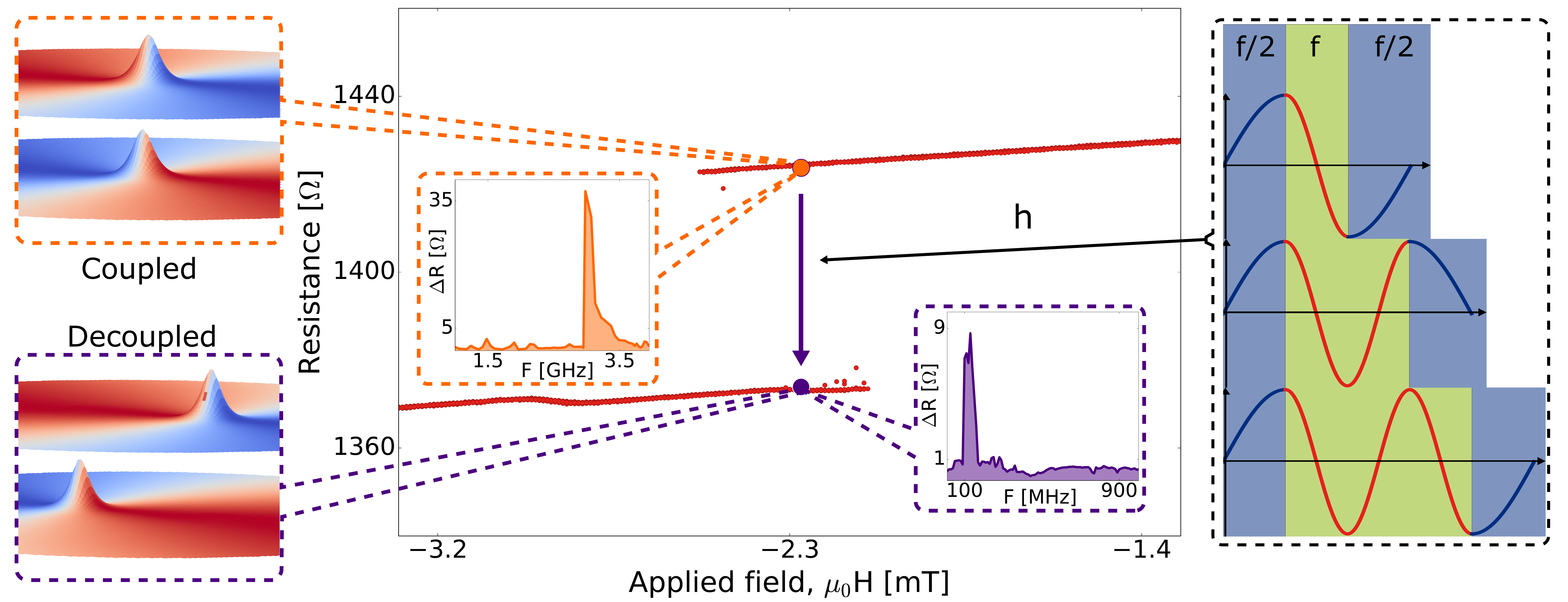}
\caption{Measured R-H decoupling-recoupling hysteresis of a representative junction at 77~K, where the resistance reflects the position of the bottom core. Prior to decoupling at about -2.4~mT, the cores are strongly bound and centered in the particle (\emph{coupled state}). At higher negative fields the cores separate, producing two essentially independent vortices (\emph{decoupled state}). Reversing the dc field brings the cores back into the coupled state at -2.2~mT. The panels on the left show a micromagnetic simulation of the spin distribution for these two states, where the height gives the out-of-plane magnetization component and the blue and red represent the positive and negative in-plane easy-axis magnetization. Typical resonant decoupling waveforms used in our experiment are shown in the right panel, where the first and last quarter-periods have half the frequency of the center part of the waveform (1/2 to 3/2 inner periods shown). Insets show the spectra in the coupled and decoupled core states.}
\label{fig1}
\end{figure*}

The samples measured in this study were elliptical nanopillars comprised of a free synthetic antiferromagnet (SAF) of two Permalloy (Py) particles, each 5 nm thick and 350 to 420 nm in plane, separated by a tantalum nitride spacer of thickness 1 nm. The spacer is chosen such as to suppress all direct and indirect exchange coupling between the Py layers. The free SAF is separated from a pinned flux-closed read-out layer by a tunnel barrier of Al-O. 50$~\Omega$ Cu lines were integrated on-chip for broadband write and readout of the junctions. Prior to vortex dynamics measurements, the Py SAF particles were set in the P-AP vortex state. The mutually parallel core polarizations result in a strong attraction between the cores, while the antiparallel chiralities make the two cores move in opposite directions under external magnetic fields. More details on the sample fabrication and measurement methods can be found elsewhere.\cite{Korenivski2005,Gaidis2006,Konovalenko2009}

In its ground state, a P-AP vortex pair is centered in the nano-particle, with the two cores strongly bound on the vertical axis at near zero lateral core separation. A micromagnetic simulation of the corresponding spin distribution is shown in the top left panel of Fig.~\ref{fig1}. Due to the antiparallel chiralities of the two vortices, an applied field of some threshold strength, via its Zeeman effect on the vortex periphery spins, overcomes the on-axis attraction of the parallel cores, causing them to decouple and move toward the opposite sides of the particles, as shown micromagnetically in the bottom left panel of Fig.~\ref{fig1}. The two states have entirely different spectral properties as shown by the insets. Without thermal agitation, this transition should be hysteretic, which is indeed observed in our experiments (main panel of Fig.~\ref{fig1}), with the lifetime of the two states of the order of hours to days for dc biasing in the center of the core-core hysteresis, and infinite lifetime of the coupled state in zero field.

Of particular interest and to date unexplored is the transient dynamics of the decoupling process, which we study by biasing P-AP vortex pairs to mid-hysteresis and exciting them with pulses of duration that are multiples of the half-period of the pair's rotational resonance frequency (2-3 GHz). The junction resistance was measured post-excitation and, had the vortex cores decoupled, the biasing dc field was toggled to reset the coupled state. The full excite-measure-reset sequence was about 100~ms per point -- a vanishing fraction of the states' lifetime. The ac field pulses were produced using a 5~GS/s arbitrary waveform generator and are shown in the right panel of Fig.~\ref{fig1}. The shortest waveforms are significantly anharmonic due to the first and last quarter periods having half the frequency of the inner waveform section.

\begin{figure}[!t]
\includegraphics[width=3.35in]{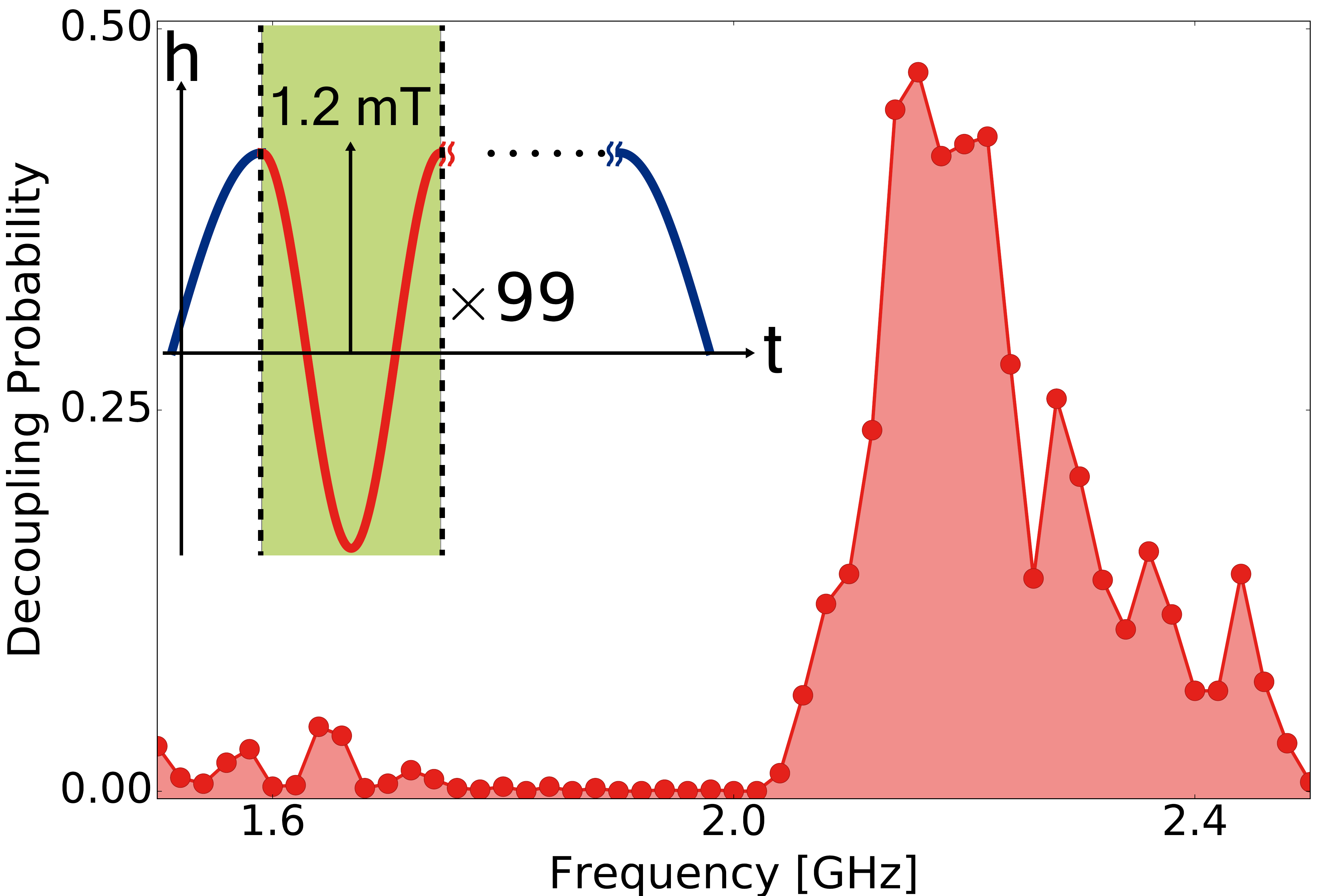}
\caption{Frequency dependence of the decoupling probability for 100 periods of ac field, measured with the ac field amplitude of 1.2~mT and dc field bias at the center of the core-core hysteresis. The anharmonic excitation waveform used is shown in the inset, where the center area (green) is a harmonic sine-wave of a given frequency, repeated 99 times. The decoupling probability shows a clear peak in the range of the rotational resonance of the P-AP pair, slightly lower in frequency than the zero-field eigenmode (inset Fig.~\ref{fig1}) due to an increased core-core separation under the applied dc field bias.}
\label{fig2}
\end{figure}

Forced core oscillations of significant amplitude only occur if the external excitation is in resonance with either the gyrational or rotational mode of the vortex pair.\cite{Cherepov2012} The frequency dependence of the core-core decoupling probability measured around the rotational resonance is plotted in Fig.~\ref{fig2} and shows a clear resonant behavior, peaked at about 2.2~GHz. The duration of the excitation used of 100 periods or about 50 nanoseconds is much longer than any intrinsic oscillation period -- sub-nanosecond for the rotational and a few nanoseconds for the gyrational modes, respectively. This combined with the ac field amplitude well below the dynamic switching threshold (the case for the data in Fig.~\ref{fig2}) represents the classical configuration of switching by thermal escape.

\begin{figure}[!t]
\includegraphics[width=3.35in]{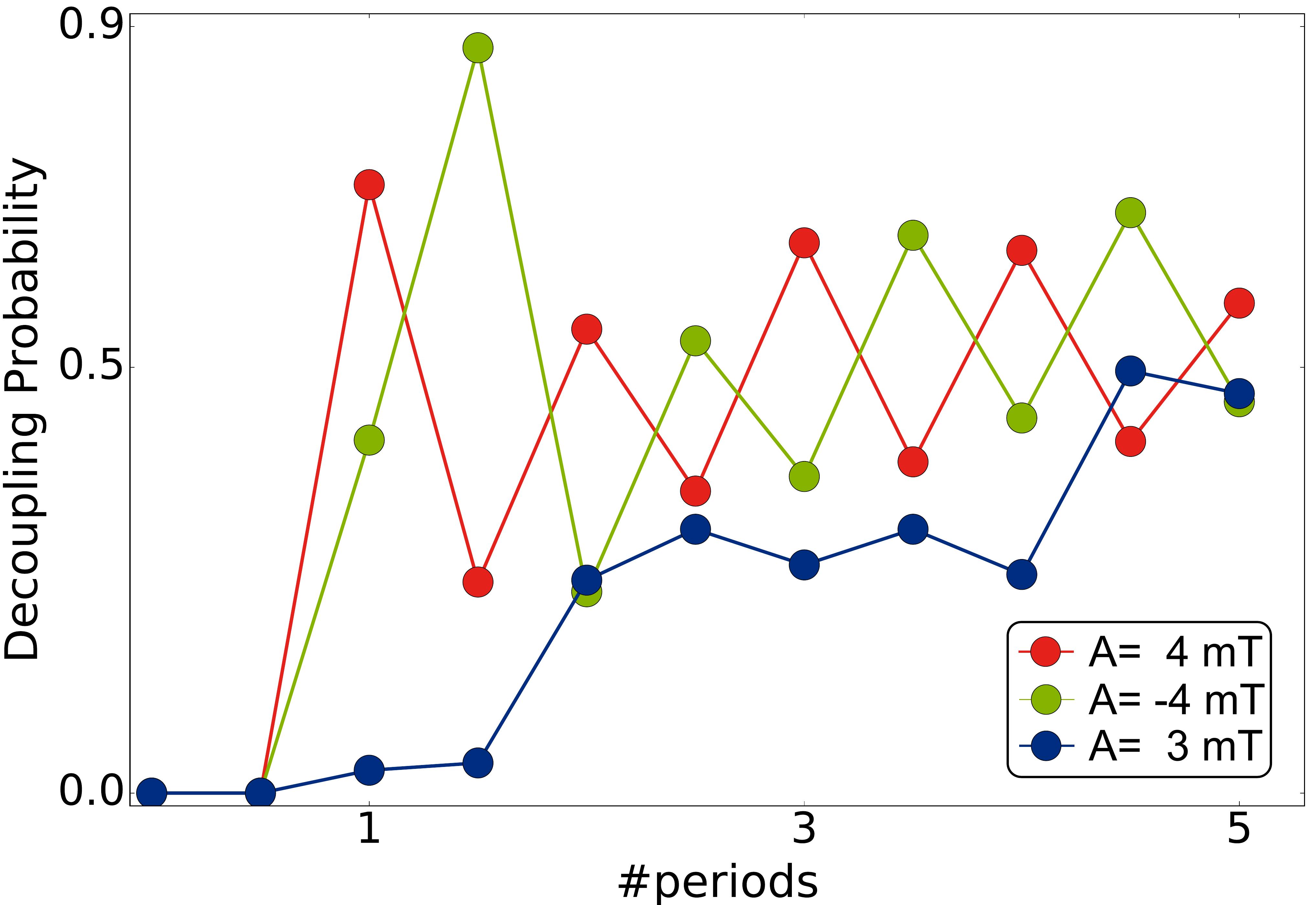}
\caption{Oscillation in the measured decoupling probability in the transient dynamic regime for the excitation duration of the order of one period of the rotational frequency (x-scale), for the rising edge of the excitation waveform parallel (red) and antiparallel (green) to the biasing dc field direction. Shown for comparison is the same decoupling probability measured at a lower field (3~mT, blue data points), corresponding to a non-dynamic regime, with switching predominantly by thermal agitation. The frequency of the inner section of the anharmonic ac field waveform is 2.2~GHz.}
\label{fig3}
\end{figure}

For short-duration high-amplitude excitation, the decoupling process is dynamic rather than stochastic in nature, with the decoupling probability strongly oscillating with the number of half periods of the ac field, as shown by the data in Fig.~\ref{fig3} (red and green). This is in contrast to a lower-amplitude thermally assited process, where the switching probability shows a nearly monotonous increase versus the excitation duration (blue). Two periods correspond to approximately 1 nanosecond total excitation time. Not shown is the asymptotic thermally-assisted behavior of the probability, where it gradually increases with the number of periods and approaches unity, while the dynamic-regime probability continues to show oscillations and levels off at about one half.

Using a model based on the Thiele equations,\cite{Thiele1973} modified with our specific core-core interaction potential, we have shown that the dynamics of a P-AP vortex pair is fully described by the core-core separation vector, $\mathbf{x}$, and the pair's offset vector, $\mathbf{X}$. Due to the symmetry of the structure, the motion of the ``center-off-mass" can be neglected even under resonant excitation. The resulting equations of motion contain the separation vector only:
\begin{eqnarray}
G \left( \mathbf{e}_{z} \times \frac{d\mathbf{x}}{dt} \right)=\frac{\partial U}{\partial \mathbf{x}} +\lambda \frac{d\mathbf{x}}{dt} +C[\mathbf{e}_{z} \times \mathbf{h}(t)],
\end{eqnarray}
where $G$ is the effective gyroconstant, $\lambda$ is the viscous friction constant, and the last term reflects the effect of the ac field. The coefficients are described in detail in Ref.\cite{Holmgren2018}. Potential $U(\mathbf{x})$ is a sum of the boundary repulsion potential, Zeeman energy due to the dc field, and the core-core interaction potential, detailed in Ref.\cite{Cherepov2012}.

In the region of hysteresis, $U(\mathbf{x})$ has two potential wells, corresponding to the coupled and decoupled states. A high frequency excitation can lift the system out of the coupled well, very efficiently if the waveform is in resonance with the eigenmode of the given state. Post excitation, the core motion is well described by the phase-plane methods,\cite{Bondarenko2017} the energy can only decrease, and the two potential wells transform into basins of attraction for the coupled and decoupled states. Even if the cores were transiently decoupled, the ballistics of the system is such that the final state of the pair can be either decoupled or, in fact, coupled. We have previously shown that the topography of the trajectories leading to the coupled/decoupled wells in phase space is quite interesting: a shift in the post-excitation position by as little as a few nanometers can change whether the pair remains decoupled or is recaptured by the coupled-state attractor.\cite{Bondarenko2017} The fraction of the phase-space covered by the trajectories focused into the decoupled state increases toward unity as the biasing field is increased toward the outer side of the core-hysteresis, and vanishes at the inner side of the hysteresis. The oscillation in the decoupling probability seen in the data of Fig.~\ref{fig3} is thus a consequence of the topography of the phase-space of a vortex pair subject to excitations of strength comparable or higher than the intrinsic core-core coupling and thermal fluctuations. Our micromagnetic simulations arrive at the same conclusions and are used below to illustrate the theory.

\begin{figure*}[!t]
\includegraphics[width=4.85in]{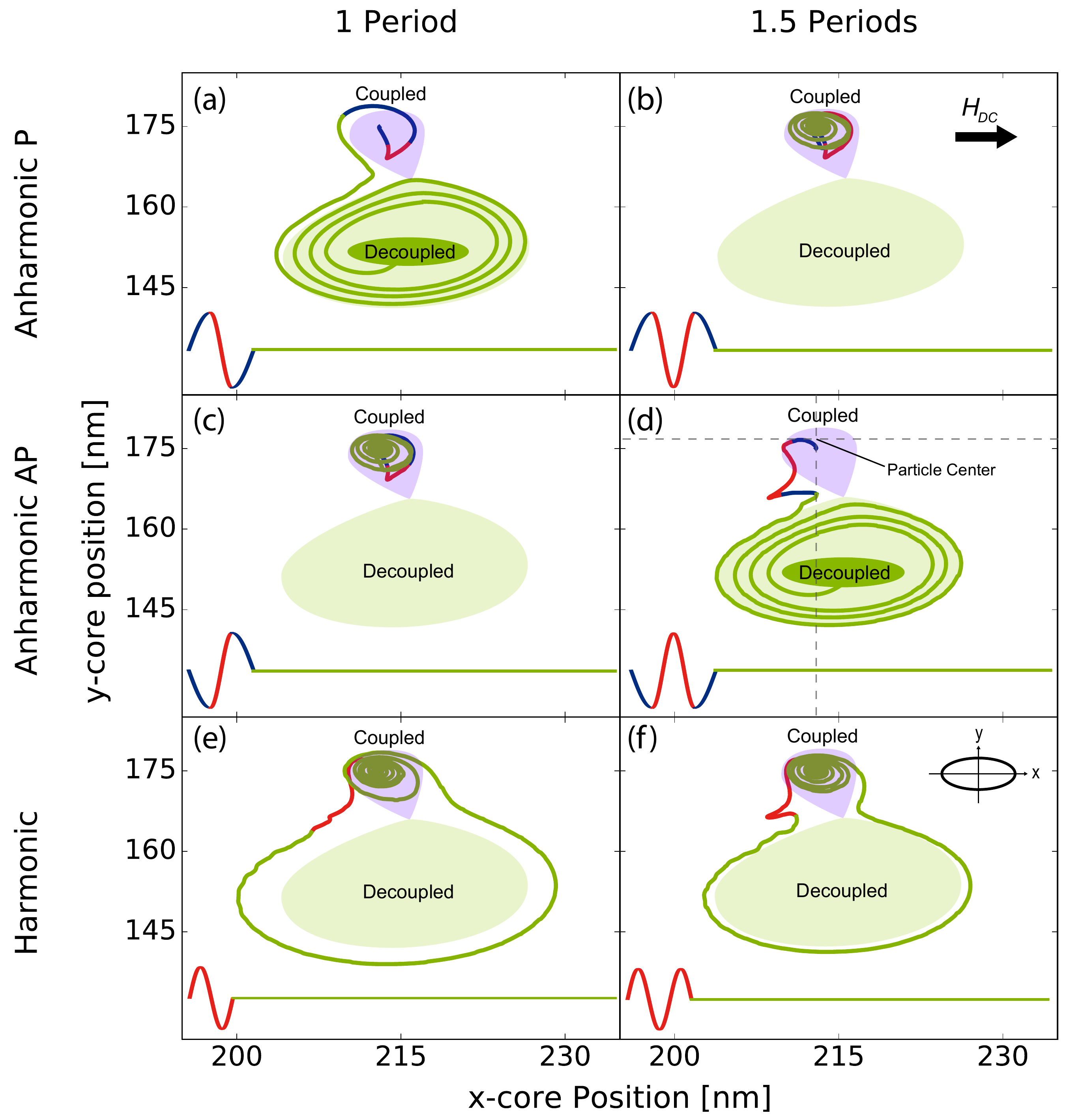}
\caption{Micromagnetically simulated characteristic trajectories of the vortex core in the bottom layer, showing the effects of resonant anharmonic excitation (2.5~mT in amplitude throughout). For (a-d) the excitation is the same as that used on the experiment (Fig.~\ref{fig3}) -- anharmonic, with the center-frequency of 2.2~GHz. In (a,b) the initial rise of the ac field is parallel to the direction of the dc biasing field; in (c,d) -- it is antiparallel. (e) and (f) show the effect of harmonic excitation at 2.2~GHz, with the initial ac field rise parallel to the bias field (reversing the sign to AP changes neither results in core-core decoupling). The left column shows the trajectory for the total excitation duration of 1 period; the right column -- 1.5 periods. The trajectories are split in three colors to indicate the corresponding section of the excitation waveform: blue sections correspond to 1.1~GHz, red -- to 2.2~GHz, and green -- to the inertial relaxation period with only the static biasing field on (zero ac field).}
\label{fig4}
\end{figure*}

Dynamic core decoupling occurs as fast as a single period of a resonant field and depends on its initial phase -- whether the excitation swings initially along or counter to the dc bias -- as seen from the anti-phase oscillations (red vs. green) in the measured switching probability in Fig.~\ref{fig3}. The same behavior is seen in our micromagnetic simulations, performed using GPU-accelerated Mumax3\cite{Vansteenkiste2014} and summarized in Fig.~\ref{fig4}. The cell size used was $\{x,y,z\}=\{1.76471,1.76471,2\}$~nm with the mesh of 240$\times$200$\times$5 cells, with the standard Py parameters ($M_s=850$~kA/m, $\alpha=0.013$, $A=13$~pJ/m) for the two particles, spaced by one empty cell in z. The magnetic layers were biased to the center of the simulated hysteresis, resonantly excited, and time-relaxed for 20 ns.

Figure~\ref{fig4} explains and illustrates the mechanisms discussed above by showing micromagnetically simulated transient core trajectories for a set of characteristic switching events. The exact waveforms used to induce these transitions are shown in the bottom of each panel. Common to all six waveforms is that the total duration of the excitation is shorter or much shorter than the time of switching where it occurs. Panels (a) versus (c) demonstrate how the initial phase of a single-period resonant excitation can determine whether the core-core state switches or not. When the ac field pulse starts along the dc bias, swings far back (promoting closer cores), and finally swings the cores apart again, the kinetics acquired by the cores is sufficient to decouple the pair inertially -- long past the excitation is turned off. In contrast, the initial swing opposite to the dc bias, followed by a forward swing provides insufficient kinetics for decoupling.

The clear resonant character of the observed transitions is evident from comparing Fig.~\ref{fig4}(a) versus (b) and (c) versus (d) -- adding one half period to the respective waveforms fully reverses the effect, such that the `parallel' initial phase leaves the core-core state unaffected while the `antiparallel' phase decouples the pair. 

Another interesting and surprisingly useful effect is that of waveform anharmonicity, illustrated by comparing Fig.~\ref{fig4}(a) versus (e) and (b) versus (f). The excitation waveforms of (e) and (f) are harmonic, whereas those of (a) and (b) have the first and last quarter-periods of half the frequency of the center section; all other parameters equal. The 'slingshot' observed in (a), with its slower final swing, is just right to put the cores into the decoupled attractor, in which they eventually settle. It is interesting to point out that the dissipation in the system is rather low, so the post-excitation kinetics often results in the cores circling fully around the decoupled-state attractor to eventually settle in the original coupled-state (e,f).

In conclusion, we have investigated the dynamics of the core-decoupling transition of a vertically stacked spin-vortex pair with strong quasi-monopole core-core attraction, when subject to resonant field excitation. Using field waveforms with frequencies around the rotational eigenmode of the vortex pair, sufficient core kinetics can be induced to decouple the pair, with the decoupling probability strongly oscillating versus added duration in multiples of the half period of the resonance frequency. The slingshot-effect of anharmonicity added to the excitation waveform can be used to efficiently navigate the cores into the desired final state and thereby reduce the number of half-periods required to core-decouple the vortex pair. We demonstrate how 100 ps range field pulses can be used for reliable switching between core-pair states with greatly different static and spectral properties. These results should be useful in designing vortex based memory and multi-frequency oscillator devices.

\section*{Acknowledgment}
Swedish Research Council (VR) grant 2014-4548, Program of NUST ``MISiS" grant No. 2-2017-005, implemented by a governmental decree dated 16th of March 2013, No 211, and the National Academy of Sciences of Ukraine Project 1/17-N are gratefully acknowledged.  

\bibliographystyle{unsrt}
\bibliography{refs}

\end{document}